\documentclass[useAMS,usenatbib]{mn2e}
\newcommand{\OIII}{\mbox{[O III]}}
\newcommand{\NII}{\mbox{[N II]}}
\newcommand{\OI}{\mbox{[O I]}}
\newcommand{\SII}{\mbox{[S II]}}
\newcommand{\HII}{\mbox{H II}}
\newcommand{\NIIHa}{\NII/H$\alpha$}
\newcommand{\SIIHa}{\SII/H$\alpha$}
\newcommand{\OIHa}{\OI/H$\alpha$}
\newcommand{\OIIIHb}{\OIII/H$\beta$}
\newcommand{\Ha}{H$\alpha$}
\newcommand{\Hb}{H$\beta$}

\usepackage{epsfig} % to have figures in the .eps format
\usepackage{epstopdf}
\usepackage{graphicx}
\usepackage{amsmath} %math stuff
\usepackage{hyperref}
\usepackage{times}
\usepackage{verbatim}
\usepackage{color}

\usepackage{breakcites}

\begin{document}

\title[Starburst-AGN mixing: I. NGC 7130]{Starburst-AGN mixing: I. NGC 7130}

\author[R. L. Davies et al.]{Rebecca L. Davies,$^1$\thanks{Email:Rebecca.Davies@anu.edu.au} Jeffrey A. Rich, $^2$ Lisa J. Kewley, $^{1,3}$ Michael A. Dopita, $^{1,4}$ \\ \\
$^1$Research School of Astronomy and Astrophysics, Australian National University, Cotter Road, Weston, ACT 2611, Australia \\
$^2$Carnegie Observatories, 813 Santa Barbara Street, Pasadena, California, 91101 USA \\
$^3$Institute for Astronomy, University of Hawaii, 2680 Woodlawn Drive, Honolulu, HI 96822 \\
$^4$Astronomy Department, King Abdulaziz University, PO Box 80203, Jeddah, Saudi Arabia}

\maketitle

\begin{abstract}
We present an integral field spectroscopic study of the Luminous Infrared Galaxy (LIRG) NGC 7130, a known starburst-AGN composite galaxy. We employ standard emission line ratio diagnostics and maps of velocity dispersion and velocity field to investigate how the dominant ionising sources change as a function of radius. From the signatures of both star formation and AGN activity we show that NGC 7130 is a remarkably clean case of starburst-AGN mixing. We find a smooth transition from AGN dominated emission in the centre to pure star forming activity further out, from which we can estimate the radius of the extended narrow line region to be 1.8 $\pm$ 0.8 kpc. We calculate that the fraction of \OIII\ luminosity due to star formation and AGN activity is 30 $\pm$ 2\% and 70 $\pm$ 3\% respectively, and that the fraction of \Ha\ luminosity due to star formation and AGN activity is 65 $\pm$ 3\% and 35 $\pm$ 2\% respectively. We conclude with a discussion of the importance and potential of starburst-AGN mixing for future studies of the starburst-AGN connection.

\end{abstract}

\section{Introduction}
\label{Sec:intro}

One of the longest standing mysteries in astrophysics is the nature of the relationship between star formation and AGN activity in composite galaxies. For many years, it has been well established that there is a strong relationship between the accretion activity of supermassive black holes and the evolution of their host galaxies \citep{Cattaneo99, Haehnelt00, Granato04, LaMassa12, Ka03, CidFernandes01}. The mass of a supermassive black hole has been found to scale with properties of the host galaxy such as velocity dispersion (M-$\sigma$ relation), the stellar mass in the bulge ($M_{BH}$-$M_{\star}$ relation) and the luminosity of the bulge ($M_{BH}$-L relation) \citep[e.g.][]{Magorrian98, Marconi03, Bennert11, McConnell13, Ferrarese00, Gebhardt00, Tremaine02, Gultekin09}. Although these relationships indicate that by z = 0, the building of the stellar mass and black hole mass in galaxies is in some way correlated, it is unclear whether such a correlation is the result of a connection between black hole and star formation activity on timescales relevant to the formation of stars (facilitated by processes such as mergers, starburst-driven winds and AGN-driven outflows, e.g. \citealt{Yuan10, Rafferty11}), or whether these scaling relations are simply pointing to something more fundamental which is yet to be uncovered \citep[e.g.][(review)]{Mullaney12,Alexander12}. 

Naively, one might suggest that the scaling relations simply fall out of the common requirement of cold gas to fuel both star-formation and AGN activity. This explanation is problematic, as gas must lose $\approx$ 99.9\% of its angular momentum in order to move from a $\approx$ 10 kpc stable orbit in a star-forming disk to the zone of influence of an AGN on a $\approx$ 10 pc scale \citep{Jogee06}. However, it is well established both theoretically and observationally that black hole activity must impact star formation. Black hole feedback can heat material in the inter-stellar medium of a galaxy \citep[e.g.][]{Sijacki06}, or drive powerful outflows resulting in significant gas redistribution \citep[e.g.][]{Alatalo11, Rich10, Rich11, DeBuhr11}. These processes deplete the galaxy's cold gas reservoir, resulting in the regulation or quenching of star formation \citep[e.g.][]{DiMatteo05, Croton06, Hopkins06, Sijacki07, Hopkins08, Booth09, McCarthy10}. Black hole feedback directly impacts the results of galaxy evolution simulations, which consistently indicate that AGN feedback is pivotal for preventing over-production of stellar mass in galaxies \citep[e.g.][]{Vogelsberger13, Croton06}. 

However, the finding that approximately 30-50\% of Seyfert 2 nuclei are associated with young stellar populations \citep[e.g.][]{Sarzi07, GonzalezDelgado05, CidFernandes04, GonzalezDelgado01, Storchi-Bergmann01} seems to suggest that AGN activity may enhance star-formation, and/or that circumnuclear star-formation may enhance AGN activity. \citet{Silverman09} found that the vast majority of AGN hosts at z $\leq$ 1 have star formation rates higher than that of non-AGN hosts of similar mass, and that the incidence of AGN activity increases with decreasing stellar age. 

All this evidence suggests that star formation itself is an important source of viscosity in the circumnuclear gas, assisting some of the gas to fall into the AGN, while driving the rest into states of higher angular momentum. In order to gain greater insight into the starburst-AGN connection, it is necessary to study the relationship between black hole and star formation activity on relatively instantaneous timescales. However, it is notoriously difficult to accurately calculate quantities such as star formation rates in starburst-AGN composite galaxies. In such galaxies, the photometric and spectroscopic indicators of star formation will be contaminated by emission from the AGN, so that it is only possible to derive upper limits on star formation rates, with an extremely large margin for error. 

Several studies have attempted to combat this issue by calculating the relative contribution of AGN and star formation in starburst-AGN composite galaxies. \citet{Heckman04} compute the AGN fractions for galaxies from the Sloan Digital Sky Survey (SDSS; \citealt{York00}) Data Release 4 (DR4) by comparing the positions of galaxies on the \NIIHa\ vs \OIIIHb\ diagnostic diagram to synthetic composite galaxy templates with different AGN fractions. \citet{Davies07} estimate the contribution of the AGN to the line emission of a sample of Seyfert galaxies by calculating the equivalent width of a CO absorption feature and comparing this to the known value for globular clusters. \citet{Imanishi11} constrain the contribution of the AGN in the infrared by analysing the strength of polycyclic aromatic hydrocarbon (PAH) features in the 3-20$\mu$m range. All of these methods assume that AGN fractions can be reliably calibrated across the AGN sequence. However, variations in ionisation parameter, metallicity, hardness of the radiation field and electron density across the active galaxy population introduce significant uncertainty in the derived AGN fractions when information from multiple galaxies is combined (e.g. \citealt{Ke13}). 

The advent of integral field spectrographs has provided astronomers with the tools required to conduct accurate and detailed studies of galaxy power sources. Nuclear spectra, such as those obtained by SDSS, can provide information regarding only the very centre of the galaxy (requiring correction for significant aperture effects, see \citealt{Ke05}). However, given several hundreds or even thousands of spectra from across the galaxy, it is possible to study the variation of the ionisation state and velocity field of gas throughout the galaxy. To date, some of the most elegant results from IFU data have related to the study of shock-starburst composite galaxies. For example, \citeauthor{Rich10} (\citeyear{Rich10,Rich11}) used IFU data to form the basis of a convincing argument for the identification of two Luminous Infrared Galaxies (LIRGs) - IC 1623 and NGC 3256 - as shock-starburst composites, and in another instance to show that the extended LINER-like emission of NGC 839 can be best described by shocks originating from a galactic superwind. \citet{Fogarty12} also used line ratio maps and velocity field information from the Sydney AAO Multi-Object IFS (SAMI) to identify a galactic wind in ESO 185-G031. 

IFU data are also of particular interest to the study of the starburst-AGN connection, as they allow for the separation of star-forming and AGN dominated regions. Already, several IFU studies, both in the optical and in the infrared, have identified and investigated starburst-AGN composite galaxies, such as NGC 1068 \citep{Gerssen06}, NGC 5135 \citep{Bedregal09}, IRAS 19254-7245 \citep{Reunanen07} and HE 2211-3903 \citep{Scharwachter11}. IFU data also provide the information required to study how the contribution of the starburst and the AGN vary with distance from the central AGN, probing the radius of the AGN Extended Narrow Line Region (ENLR). 

In this paper, we capitalise on the the strengths of IFU data to carry out a detailed study of the ionisation sources and kinematics in NGC 7130. NGC 7130 is a LIRG, observed using the Wide Field Spectrograph (WiFeS) on the ANU 2.3m telescope as part of the Great Observatory All-Sky LIRG Survey (GOALS; \citealt{Armus09}). LIRGs are particularly appropriate targets for study of the starburst-AGN connection as they commonly contain both starbursts and AGNs. The prevalence of dual energy sources in LIRGs is likely to be a direct result of their history as mergers of molecular gas-rich spirals, which are thought to be the trigger for powerful bursts of star formation (e.g \citealt{Sanders96, Barnes92}). During merger events, gaseous dissipation funnels material towards the centre of mass of coalescing systems, leaving them with extremely large reservoirs of gas. This gas rotates in the gravitational field of the galaxy, and can be converted into a rotating stellar disk by means of a starburst (e.g. \citealt{Barnes92}, \citeauthor{Mihos94} \citeyear{Mihos94,Mihos96}) and/or feed AGN accretion processes.

Evidence for the starburst-AGN composite nature of NGC 7130 has been found in the infrared, optical, UV and X-ray (\citealt{Spinoglio02, Levenson02, Contini02, GonzalezDelgado98, CidFernandes01, GonzalezDelgado01}). In this paper, we use this galaxy to show that IFU data can be utilised to create spatial maps which allow a clean separation of regions of star formation, AGN and composite ionisation from both sources, and permit the identification of outflows and/or feedback that may be present in the galaxy. We demonstrate that there is a strong relationship between the ionisation state of the gas and distance from the centre of the galaxy - a clear indicator of starburst-AGN mixing. We explore how well IFU data can resolve the size of the AGN extended narrow line region and the overall contribution of the starburst and the AGN to optical line emission in individual spaxels.

We summarize the existing observational framework for the composite starburst-AGN nature of NGC 7130 in Section \ref{Sec:Sec2}. We discuss our observations and data reduction in Section \ref{Sec:Sec3} and present emission line maps in Section \ref{Sec:Sec4}. In Section \ref{Sec:Sec5} we present emission line diagnostic diagrams, discuss indicators of starburst-AGN mixing and quantify the size of the AGN ENLR, calculate starburst and AGN fractions and investigate how these fractions vary as a function of radius. We analyse velocity field information in Section \ref{Sec:Sec6}, finding further support for the composite nature of this galaxy. We discuss the implications of our findings in Section \ref{Sec:Sec7} and present our conclusions in Section \ref{Sec:Sec8}. 

\begin{figure}
\centerline{\includegraphics[scale=0.65]{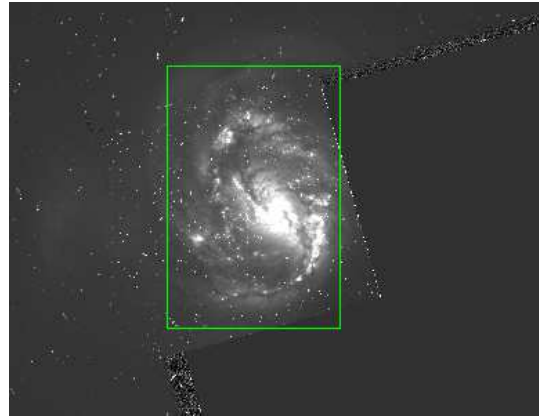}}
\caption{WiFeS Pointing (solid green rectangle) overlaid on a WFPC2 image of NGC 7130 taken with the \emph{Hubble Space Telescope} (F606W, 6030$\AA$; \citealt{Malkan98}).}\label{Fig1}
\end{figure} 

\begin{figure*}
\centerline{\includegraphics[scale=0.7]{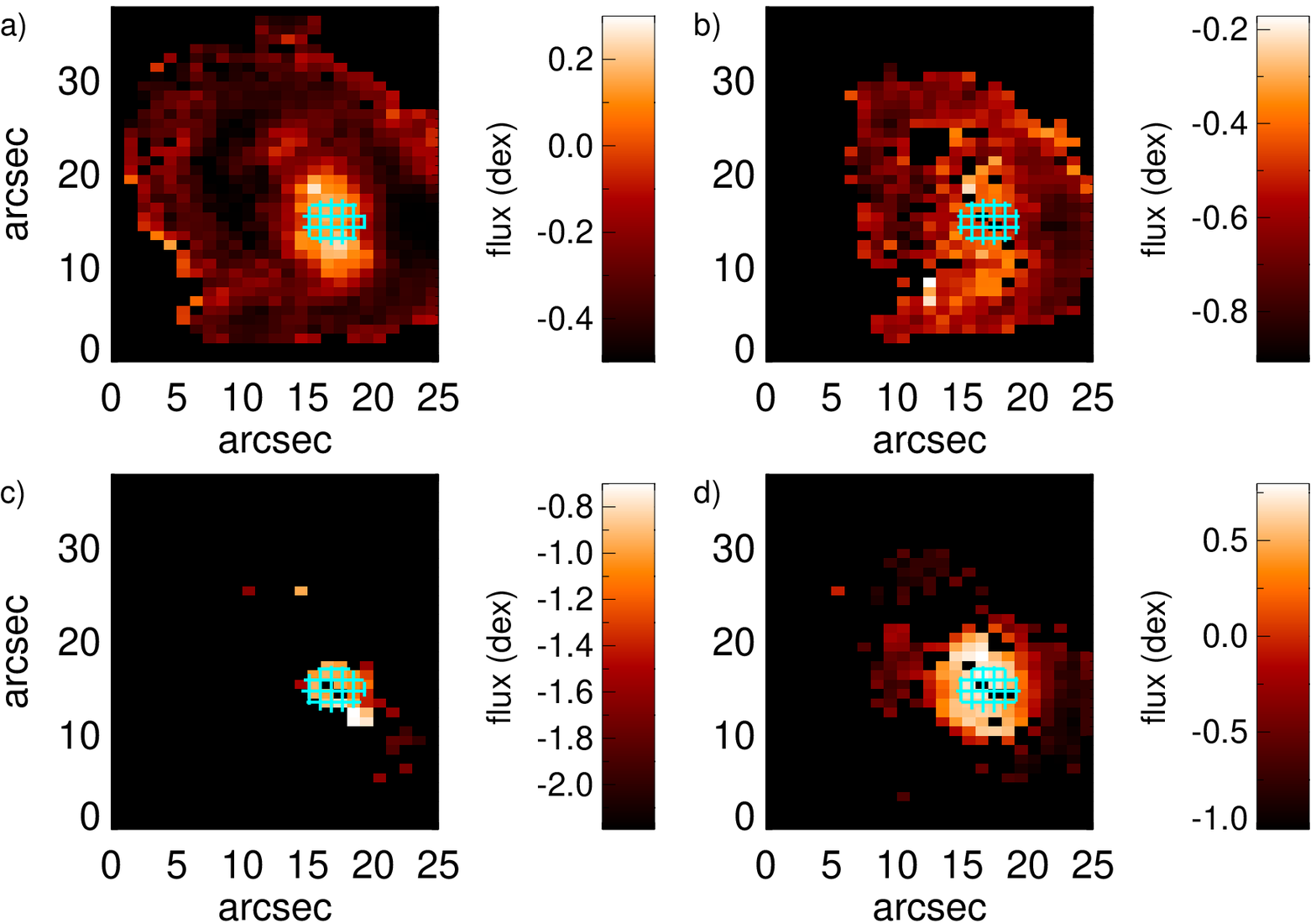}}
\caption{Top to bottom, left to right: Maps of (a) \NII/H$\alpha$, (b) \SII/H$\alpha$, (c) \OI/H$\alpha$, and (d) \OIII/H$\beta$. The maps have dimensions $25''\times38''$, with the same orientation as the HST image in Figure \ref{Fig1}. Cyan crosshairs illustrate the location of the AGN (as identified by the \OIHa\ peak) on all line ratio maps. The nuclear region of the galaxy clearly has enhanced line ratios, indicating the presence of an AGN. A ring of star-formation is also visible in the \NIIHa\ map. 
}\label{Fig3}
\end{figure*} 

Throughout this paper we adopt cosmological parameters $H_{0} = 70.5{\rm kms}^{-1}{\rm Mpc}^{-1}$, ${\Omega}_{\Lambda} = 0.73$, and $\Omega_{M}=0.27$ based on the 5-year WMAP results by \citet{Hinshaw09} and consistent with flat $\Lambda$-CDM cosmology. 

\section{NGC 7130}
\label{Sec:Sec2}

NGC 7130 is a LIRG at a redshift of z = 0.016 (corresponding to a spatial scale of $\sim$ 340 pc $\rm arcsec^{-1}$), with log($L_{IR}$/$L_{\odot}$) = 11.42 \citep{Armus09}. The galaxy is located at RA = $21^{h}48^{m}19.49^{s}$ and DEC = $-34^\circ$57'04.7". It is a peculiar spiral galaxy \citep{Corwin94} with two close dwarf galaxy companions \citep{GonzalezDelgado98}, and hosts a Seyfert 1.9 nucleus \citep{Veron-Cetty06} surrounded by a powerful and compact (90 pc) circumnuclear starburst \citep{Levenson02}. Parts of the galaxy are strongly obscured - in particular the circumnuclear region, in which the typical obscuration along the line of sight is Compton thick ($N_{H} >1.5\times10^{24}\, \rm cm^{-2}$) (\citealt{Levenson02}). However, the nuclear region is not as obscured in other directions, allowing ionizing photons to escape and excite the galaxy's ENLR.  

NGC 7130 has previously been studied in the IR, optical, UV, and X-ray. The predominately young stellar population can be detected strongly in the far infrared (25-500$\mu$m) (\citealt{GonzalezDelgado01}), and the majority of the near ($\approx$ 0.8-2.5$\mu$m) and mid (2.5-25$\mu$m) infrared emission can be explained by thermal emission of dust shock-heated by supernova ejecta and stellar winds (\citealt{Contini02}). The blue optical color of the galactic nucleus and the presence of clear Balmer and neutral helium absorption lines are also characteristic of young and intermediate age stars (\citealt{GonzalezDelgado01}). The UV emission of NGC 7130 is dominated by the ``featureless continuum'' of the young stellar population, which displays clear absorption lines formed in the photospheres of O-B stars. Stellar wind resonance lines are also identified, which again are tracers of a young stellar population (\citealt{GonzalezDelgado98}). Finally, the presence of young stars is clear from the extended soft, thermal X-ray emission as well as some hard X-ray emission from X-ray binaries and other ultra-luminous X-ray sources (\citealt{Levenson02}).

Several indicators of AGN activity have been identified across a broad wavelength range. In the infrared, \mbox{Ne V} (\citealt{Petric11}) and elevated \mbox{[Ne III]} 15.56$\mu$m/\mbox{[Ne II]} 12.81$\mu$m ratios (\citealt{Pereira-Santaella10, Diaz-Santos10}) are indicators of a hard ionising radiation field,  whilst nuclear \mbox{[Si IV]} 10.51$\mu$m emission and weak nuclear PAH emission serve as indicators of AGN accretion activity (\citealt{Pereira-Santaella10, Diaz-Santos10,Stierwalt13}). The contribution of the AGN to the radiation field of NGC 7130 is also evident in the hard X-ray field, which is compact and strongly concentrated in the nuclear region of the galaxy (\citealt{Levenson02}).

It is interesting to note that both optical and X-ray studies of NGC 7130 find evidence in support of an outflow from the centre of the galaxy. \citet{Veilleux95} identify distinct velocity components in the \OIII\ and $H\beta$ emission lines, characteristic of a galactic wind attributed to the starburst activity in the circumnuclear region. A similar analysis by \citet{Levenson02} leads to the conclusion that a spectrally soft outflowing wind is present in the galaxy. \citet{Bellocchi12} also identify an outflow in the inner regions of the galaxy based on the presence of broad, blue-shifted optical emission-line profiles.

\section{Observation and Data Reduction}
\label{Sec:Sec3}

\begin{figure*}
\centerline{\includegraphics[scale=0.85]{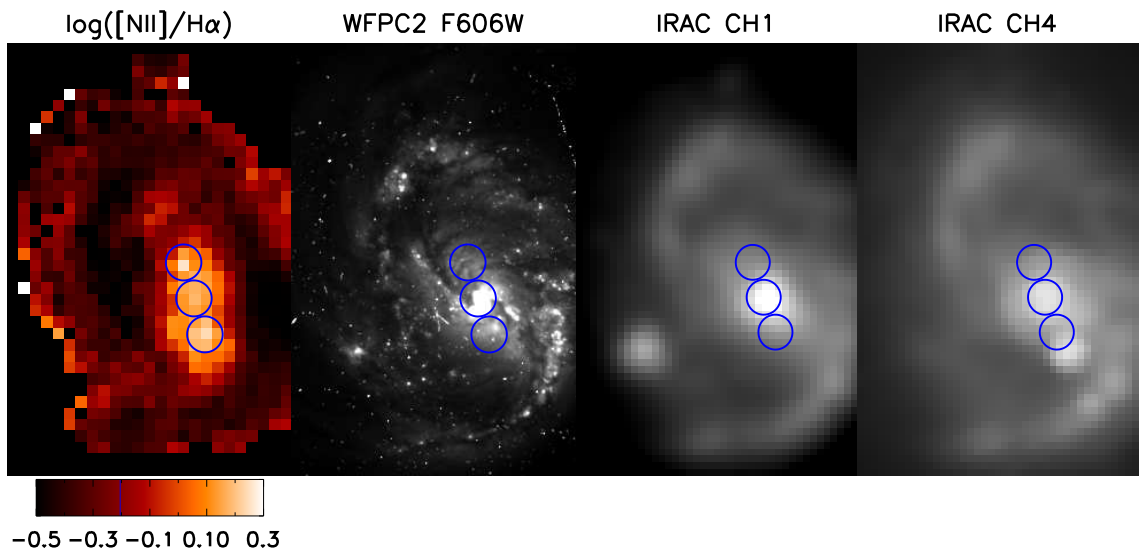}}
\caption{Left to right: \NII/H$\alpha$ map (as in Figure 2) with three blue circles overlaid on the regions where the \NIIHa\ ratio peaks, WFPC2 660W \emph{Hubble Space Telescope} image, IRAC Channel 1  (3.6$\mu$m) image, IRAC Channel 4 (8$\mu$m) image. The physical scale, coverage on the galaxy, orientation of the images and location of the blue circles remain consistent between panels. In all of the last three images it is clear than the galaxy has only one active nucleus. Both the IRAC channels also show evidence of merger activity in the form of streamers and a trailing spiral arm.}\label{Fig5}
\end{figure*}

The data used in this paper were taken using the WiFeS instrument (\citealt{Dopita07}, \citeyear{Dopita10}) on the ANU 2.3m telescope at Siding Spring Observatory. WiFeS is an integral field spectrograph which can provide a full spectral coverage through optical wavelengths over a field of view of 25'' $\times$ 38''. It is a double-beam instrument providing optimised throughput at resolutions of either 3000 or 7000 throughout the spectral coverage. In the data presented here the red arm (5700 - 7000$\AA$) was observed with the R = 7000 grating, and the blue spectrum (3700 - 5700$\AA$) was observed with the R = 3000 grating. The coverage and orientation of the WiFeS pointing relative to a WFPC2 image taken with the \emph{Hubble Space Telescope} are shown in Figure \ref{Fig1}.

Our observations consist of three 20 minute exposures of the same pointing, taken 2009 September 18. The data were reduced using the IRAF WiFeS pipeline (\citealt{Dopita10}, see \citealt{Rich10}, \citeyear{Rich11} for a detailed description of the process). Each pointing was bias-subtracted using bias frames to account for variations in the bias levels. Quartz lamp exposures were used to flat-field the observations, while twilight sky observations were used to account for variations in illumination along the slitlets. Spatial calibration along the slit was calculated by illuminating a wire with the quartz lamp. Wavelength calibration was calculated for each frame using a NeAr arc lamp spectrum. 

The individual slitlets were combined into individual blue and red data cubes. Each data cube was flux calibrated using the flux standard EG-131, accounting for the effects of atmospheric dispersion. The two individual data cubes were median-combined to a final data cube on a common grid. The final data cube was aligned astrometrically by comparing an image generated from the WiFeS red data cube with the \emph{Hubble Space Telescope} WFPC2 F606W image (see Figure \ref{Fig1}).

From the final data cube, we extract the \Ha, \Hb, \mbox{\NII\ $\lambda$ 6583}, \mbox{\SII\ $\lambda\lambda$ 6717, 6731}, \mbox{\OIII\ $\lambda$ 5007} and \mbox{\OI\ $\lambda$ 6300} emission lines. For each emission line, we accept only spaxels in which the signal-to-noise ratio (S/N) of that line is greater than 5 to allow accurate separation and classification of different ionising sources for each spaxel across the galaxy.

The Hyperleda Database lists the inclination of NGC 7130 to the line of sight as $33.7^\circ$. Given that the majority of our analysis is based on line ratios rather than absolute flux values, the only properties impacted by inclination that we are concerned with are the size of the AGN extended narrow line region and the velocity field (see Sections \ref{Subsec:Sec52} and \ref{Sec:Sec6}).

\section{Emission line Maps}
\label{Sec:Sec4}

\begin{figure*}
\centerline{\includegraphics*[scale=0.95,clip=true,trim=0 170 0 0]{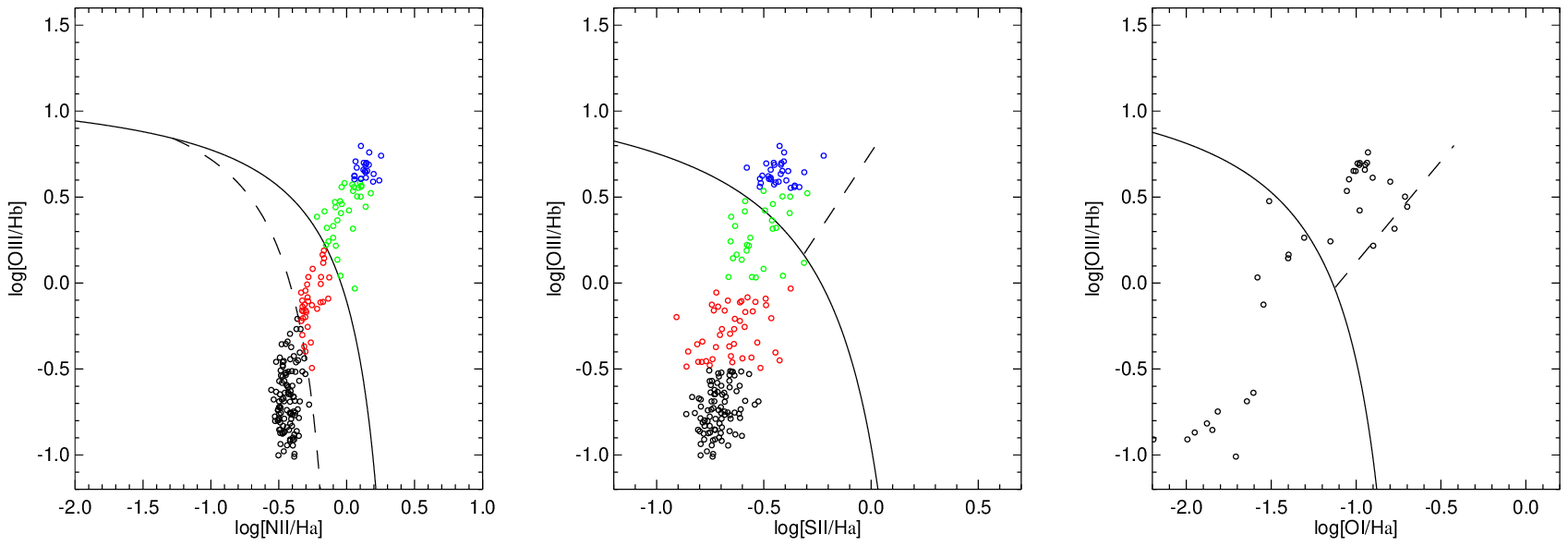}}
\caption{\NIIHa, \SIIHa\ and \OIHa\ diagnostic diagrams showing the variation in ionisation state of the gas in NGC 7130. The solid lines in the diagrams trace the theoretical upper bound to pure star-formation \citep{Ke01}, whilst the dashed line in the \NIIHa\ diagnostic diagram indicates the empirical upper bound to pure star-formation \citep{Ka03}. The smooth distribution of points from the pure star-forming region to the AGN region on the diagnostics is indicative of starburst-AGN mixing. The color-coding of points in the \NIIHa\ and \SIIHa\ diagnostic diagrams relates to the maps shown in Figure \ref{Fig7}.}\label{Fig2}
\end{figure*}

\subsection{Spatial Distribution of starburst vs. AGN}
\label{Subsec:Sec41}
 
The \NIIHa, \SIIHa, \OIHa\ and \OIIIHb\ line ratios are commonly used as diagnostics of the dominant ionisation sources in galaxies, as their differing and complementary sensitivities allow them to effectively separate starburst and AGN activity (see Section \ref{Subsec:Sec51}).

The \OIIIHb\ line ratio is primarily sensitive to the ionisation parameter and temperature of the line-emitting gas, and is enhanced both in the presence of low metallicity stars, where a lack of metals prevents the nebula from cooling efficiently, and in the presence of an AGN due to the hard EUV ionising radiation field which increases the ionisation state of the gas. This degeneracy between pure star-formation and AGN activity can only be broken by including another line ratio in the analysis. 

In star-forming galaxies, the \NIIHa\ ratio is most sensitive to metallicity, as nitrogen changes from a primary to a secondary nucleosynthetic element as metallicity rises \citep{KD02}. However, even the smallest contribution from the hard EUV radiation field of an AGN causes this ratio to become enhanced, which therefore effectively separates spectra consistent with pure star-formation from those with some contribution from an AGN. 

The \OIHa\ and \SIIHa\ ratios are also enhanced in the presence of a hard ionising radiation field due to the presence of a large partially ionised zone where \OI\ and \SII\ are collisionally excited. 

In this section, we analyse the spatial variation of the \NIIHa, \SIIHa, \OIHa\ and \OIIIHb\ line ratios in NGC 7130. The maps of all four line ratios can be found in Figure \ref{Fig3}. On each map, we overlay a blue circle enclosing the region where the \OIHa\ ratios are the highest.

The \NII/H$\alpha$ line ratio peaks in the nuclear region of the galaxy. A ring of intermediate line ratios is also visible in the outer regions of the galaxy (see Figure \ref{Fig3} a). The \SIIHa\ map does not reflect any of the features of the \NIIHa\ map, contrary to expectation - however, it is likely that the lower S/N on the \SII\ emission line can account for this discrepancy (see Figure \ref{Fig3} b). The location of the \OI/H$\alpha$ peak in Figure \ref{Fig3} c reveals the location of the AGN where the ionising radiation field is hardest. This is also consistent with the location of the AGN as traced by the infrared \citep{Spinoglio02} and X-ray \citep{Levenson02} maps. 

The \OIIIHb\ map shows similar morphology to the \NIIHa\ map but does not trace the outer ring of intermediate line ratios seen in the \NIIHa\ map (see Figure \ref{Fig3} d), indicating that the ionisation parameter values of the spaxels in this ring are significantly lower than the ionisation parameter values of the spaxels in the nuclear region. Therefore, the ring is likely to be photoionised by high metallicity star formation. It is interesting to note that both the \NIIHa\ and \OIIIHb\ maps have peaks which are much more extended than the \OIHa\ peak. The \OIHa\ peak is coincident with the central regions of the elongated peaks in the \NIIHa\ and \OIIIHb\ maps. Furthermore, the \NIIHa\ map seems to have two distinct off-nuclear peaks, suggesting that ionisation sources other than star-formation and AGN activity may be important in these regions (see Section \ref{Subsec:Sec42}). 

\subsection{Outflowing Wind}
\label{Subsec:Sec42}

We include data from the \emph{Hubble Space Telescope} WPFC2 (660W) and the Infrared Array Camera (IRAC) Channels 1 (3.6$\mu$m) and 4 (8$\mu$m) (see Figure \ref{Fig5}). These ancillary data clearly indicate that the peak of the galaxy's total luminosity occurs in the centre of the nuclear region as traced by the \NII/H$\alpha$, \OIHa\ and \OIIIHb\ maps, confirming that this is the location of the central AGN. 

However, perhaps more interesting is the information that these images provide regarding the nature of the two off-nuclear knots seen in the \NII/H$\alpha$ map. The HST and IRAC images give no indication of a second nucleus of any kind coincident with either the northern or southern off-nuclear peak, and thus these peaks cannot be ionised by either stellar or AGN photoionisation, as both sources would be prominent in the near and mid-infrared. Furthermore, the Spitzer IRS maps published in \citet{Pereira-Santaella10} show no evidence for strong off-nuclear emission. Thus, we suggest that the off-nuclear \NII/H$\alpha$ peaks are best described by an outflowing wind  shock-heating the interstellar medium (see Section \ref{Sec:Sec7} for further discussion). 

It is interesting to note the presence of a bright source above the top-most blue ring in the 8$\mu$m image. Since this channel covers 7.7$\mu$m emission, this peak may indicate intense 7.7$\mu$m PAH emission and thus trace a location of intense star-formation. However, without 24$\mu$m images we cannot confirm this hypothesis.

It should also be noted that the asymmetric optical morphology of the galaxy as well as the streamers and trailing spiral arms seen in the images from Channels 1 and 4 of IRAC (see Figure \ref{Fig5}) support the notion that this galaxy may have recently undergone merger activity. 

\section{The Starburst-AGN mixing sequence}
\label{Sec:Sec5}

\subsection{Diagnostic Diagrams}
\label{Subsec:Sec51}

Our analysis of the starburst-AGN mixing in NGC 7130 is based on strong-line ratio diagnostic diagrams. \citet{BPT81} first combined the \NII/H$\alpha$ and \OIII/H$\beta$ line ratios to form a diagnostic for the dominant emission sources of galaxies. This diagnostic was subsequently revised by \citet{Osterbrock85}, and in 1987 \citeauthor{Veilleux87} added two more diagnostics to the set, combining the \OIIIHb\ line ratio with the \SII/H$\alpha$ and \OI/H$\alpha$ line ratios respectively. 

\begin{figure*}
\centerline{\includegraphics[scale=0.9,clip=true,trim=0 175 0 0]{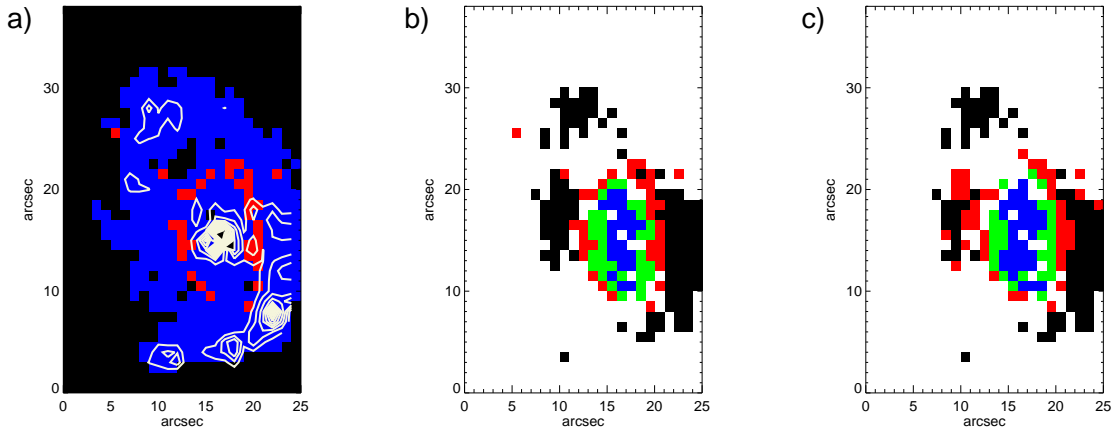}}
\caption{(a) Map of the galaxy in blue with composite spaxels plotted over the top in red. The beige contours show the large-scale H$\alpha$ morphology of the galaxy. It is clear that the composite spaxels form a ring around the nuclear region of the galaxy.(b) Map of the galaxy with spaxels color-coded according to distance up the mixing sequence on the \NIIHa\ vs \OIIIHb\ diagnostic diagram. c) Map of the galaxy with spaxels color-coded according to distance up the mixing sequence on the \SIIHa\ vs \OIIIHb\ diagnostic diagram. In panels b) and c), there is a clear relationship between the position of spaxels on the mixing sequence and their distance from the centre of the galaxy, with the contribution of the AGN decreasing towards larger radii - indicative of starburst-AGN mixing.
}\label{Fig7}
\end{figure*}

Since their original formulation, these diagnostics have been updated a series of times. \citep{Ke01} employed stellar population synthesis and photoionisation models to fit a theoretical ``maximum starburst'' (Ke01) line to all diagrams (appearing as solid lines in Figure \ref{Fig2}). These lines represent the maximum line ratios which can be attributed purely to star-formation whilst remaining consistent with theoretical models. \citet{Ka03} fitted an empirical (Ka03) line on the \NII/H$\alpha$ diagram enclosing the region where starburst-AGN composite galaxies should lie, based on the positions of the star-forming and starburst-AGN mixing sequences for a sample of over 100,000 galaxies from SDSS (\citealt{York00}). The empirical line appears dashed on the \NIIHa\ diagnostic diagram.  

Figure \ref{Fig2} shows the \NIIHa, \SIIHa\ and \OIHa\ vs \OIIIHb\ diagnostic diagrams for NGC 7130, with each data point representing an individual spaxel. In all three of the diagnostic diagrams there is a smooth distribution of points from the star-forming to the AGN region. This distribution is known as a mixing sequence, and traces variations in AGN fraction and metallicity from the HII regions outside of the AGN ENLR to the AGN in the centre of the galaxy \citep[see][]{Ke13}.

\subsection{Correlation between location and energy of ionising radiation}
\label{Subsec:Sec52}
In light of the starburst-AGN composite nature of NGC 7130, we dissect the starburst-AGN activity. The mixing sequence between the \HII\ region and AGN regimes on the 
\NIIHa\ vs \OIIIHb\ diagnostic diagram traces variations in both metallicity and AGN fraction throughout the galaxy. The contribution of the AGN is expected to rise towards the centre of the galaxy, and the metallicity may stay constant or increase towards the centre of the galaxy  (consistent with observations of metallicity gradients in local spiral galaxies; e.g \citealt{McCall82, Zaritsky94, vanZee98}). The impact of abundance gradients on our interpretation of starburst-AGN mixing will be discussed further in Section \ref{Sec:Sec7}.

In order to determine the relationship between distance from the centre of the galaxy and position on the mixing sequence, we construct maps of the galaxy, color-coding spaxels according to their location on the \NIIHa\ vs \OIIIHb\ diagnostic diagram. The results are remarkable. 

In Figure \ref{Fig7} a) we highlight the position of composite spaxels (red) relative to the rest of the galaxy (blue) and the H$\alpha$ emission peaks (beige contours). Composite spaxels are defined as those lying in between the Ke01 and Ka03 demarcation lines on the \NIIHa\ vs \OIIIHb\ diagnostic diagram, and are likely to be ionised by either a roughly even combination of star-formation and AGN activity \citep[e.g.][]{Hill06, Panessa05} or a combination of star formation and shock excitation \citep[e.g.][]{Rich11, Sharp10, Monreal-Ibero06, Farage10}. In NGC 7130, the composite points form a clean ring around the nuclear region of the galaxy. This ring is evidence of a starburst-AGN mixing region in which starburst and AGN emission are both energetically significant. Therefore, in this galaxy, the composite region demarcates the boundary of the AGN ENLR.

In Figure \ref{Fig7} b), we further separate the \NIIHa\ vs \OIIIHb\ mixing sequence into spaxels classified as \HII\ regions (black), composite (red) and AGN-dominated (blue and green). The spaxels colored blue and green occupy distinct regions on the mixing sequence (as seen in Figure \ref{Fig2}, left), with the exact division chosen so that the majority of the blue spaxels lie in the nuclear region of the galaxy. We see from Figure \ref{Fig7} b) that the correspondence between location on the mixing sequence and distance from the centre of the galaxy seen for the composite spaxels extends to spaxels with both higher and lower \OIIIHb\ values. 

The spaxels colored blue occupy the nuclear region of the galaxy, consistent with emission due to pure AGN as well as photoionisation by a galactic wind. The spaxels colored green form a clean ring around the nuclear region of the galaxy, consistent with emission dominated by AGN photoionisation with some contribution from star formation. This ring forms the inner edge of the ring observed in Figure \ref{Fig7} a), consistent with the notion that the composite spaxels are ionised by an approximately equal contribution from star-formation and AGN activity. The AGN ENLR is cleanly segregated from the lower ionisation state \HII\ region spaxels, which occupy the outer regions of the galaxy's \OIIIHb\ disk.

Figure \ref{Fig7} c) shows that very similar behaviour is uncovered when color-coding spaxels according to their distance along the \SIIHa\ vs \OIIIHb\ mixing sequence. Although the \SIIHa\ diagnostic diagram does not provide a division between pure star-forming and composite spaxels, we again observe clean rings of gas ionized by a decreasing fraction of AGN activity as distance from the centre of the galaxy increases. The spaxels with the highest \OIIIHb\ values (at the top of the starburst-AGN mixing sequence) lie closest to the centre of the galaxy, whilst the spaxels with the lowest \OIIIHb\ values lie furthest from the centre of the galaxy. It is likely that starburst-AGN mixing signatures will also be visible from the \OIHa\ diagnostic when \OI\ $\lambda$ 6300 is detected at a statistically significant level in a sufficient number of spaxels.

We use the clean segregation of the AGN ENLR to calculate its size. Assuming the ENLR to be elliptical in shape, we find that the radius of the ENLR is 5.4 $\pm$ 2.3", which corresponds to a physical size of 1.8 $\pm$ 0.8 kpc. The error on the radius encompasses the variation in length between the major and minor axes of the ellipse as well as the 2'' point-spread function (PSF) of the data. The ENLR size is consistent with the majority of ENLRs in Seyfert 2 galaxies, which are found to be 1-5 kpc in radius \citep{Bennert06}.

\subsection{Mixing sequence and ${d}_{SF}$}
\label{Subsec:Sec53}
The existence of a strong correlation between position on the mixing sequence and distance from the centre of the galaxy allows us to calculate the fractional contribution of the starburst and the AGN activity to the line emission in each spaxel. Numerical calibration of the mixing sequence is done empirically according to the notion of star-forming distance ($d_{SF}$, see \citealt{Ke06} for further discussion). The fundamental principle of $d_{SF}$ is to provide a means of quantifying the distance of a point on the \NIIHa\ vs \OIIIHb\ diagnostic diagram from the star-forming sequence, traced out by pure star-forming galaxies from SDSS \citep[see][] {Ka03}. Since the majority of AGN-dominated galaxies appear to mix with high-metallicity \HII\ regions \citep[see][]{Groves06}, $d_{SF}$ is defined such that the high metallicity end of the star-forming sequence represents $d_{SF}$ = 0.  

In global or nuclear spectra, galaxies that lie further away from the star-forming sequence can be expected to have larger AGN contributions than galaxies which lie closer to the star-forming sequence as the rate of enhancement of \OIIIHb\ scales with the collisional excitation rate in the nebula. We emphasise that this measure cannot be used to derive quantitative information regarding the absolute contribution of starburst or AGN activity to the optical luminosity of a galaxy or region when the spectra used to construct the mixing sequence are taken from several different host galaxies. Metallicity, ionisation parameter, the hardness of the AGN radiation field and electron density can all vary between galaxies and this leads to significant uncertainty in the AGN contribution \citep[see][]{Ke13}. Therefore, it is only when integral field data are available and a mixing sequence can be constructed with spaxels from a single host galaxy that the following analysis is valid.

\begin{figure}
\centerline{\includegraphics[scale=0.5]{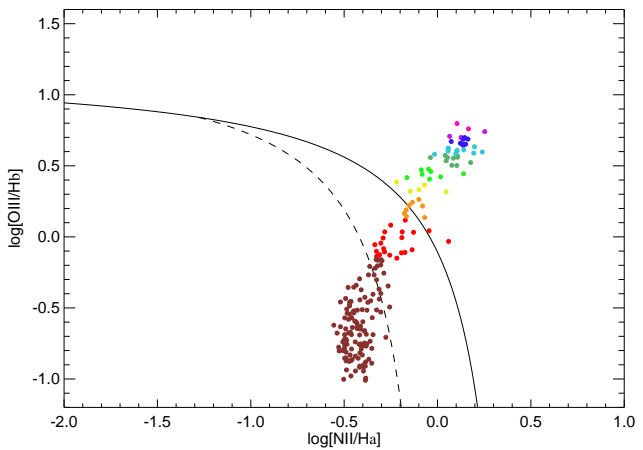}}
\caption{\NIIHa\ vs \OIIIHb\ diagnostic diagram for NGC 7130. Colors represent 10\% increments in the AGN fraction, going from 0-10\% (maroon) to 90-100\% (pink). 
}\label{Fig8}
\end{figure}

We assign 100\% and 0\% AGN fractions to the spaxels with the highest and lowest \OIIIHb\ values on the mixing sequence respectively. We then assign AGN fractions to all other spaxels according to their projection distance along the line between these two most extreme points. Figure \ref{Fig8} shows the mixing sequence for NGC 7130 color-coded by assigned AGN fractions in bins of width 10\%.

It is difficult to determine the error on the assigned AGN fractions. However, by constructing starburst-AGN mixing models we determine that the relative error in AGN fractions assigned using this technique is on the order of 6\% (see Paper II for a more detailed discussion of mixing models and error calculations).

\subsection{Calculation of the relative starburst-AGN fraction}
\label{Subsec:Sec54}

We use the numerical calibration of our starburst-AGN mixing sequence to calculate the relative contribution of the starburst and the AGN activity to the H$\alpha$ and \OIII\ luminosities of the galaxy. H$\alpha$ traces active star-formation, and thus the \Ha\ luminosity of the galaxy is expected to be largely attributable to star-formation activity. The \OIII\ emission line, on the other hand, is significantly enhanced in the presence of an AGN due to the hardness of the ionising radiation field.

The total H$\alpha$ or \OIII\ luminosity of the galaxy can be calculated using

\begin{center}
${L}_{tot}$ = $\sum_{i=1}^{n} L_n$
\end{center}

(where ${L}_{n}$ is the luminosity of the relevant line in spaxel n). The luminosity that can be attributed to the AGN is given by

\begin{center}
${L}_{AGN}$ = $\sum_{i=1}^{n} {f}^{AGN}_{n}{L}_{n}$
\end{center}

(where ${f}^{AGN}_{n}$ is the AGN fraction in spaxel n). Then the total fraction of emission due to the AGN for that emission line is given by 

\begin{center}
${f}^{AGN}_{tot}=\frac{L_{AGN}}{L_{tot}}$
\end{center}

Similarly, the total fraction of emission due to star-formation for a particular line is given by 

\begin{center}
${f}^{SF}_{tot}=\frac{L_{SF}}{L_{tot}}$
\end{center}

We find that the fraction of \OIII\ luminosity from star-formation and AGN activity is 30 $\pm$ 2\% and 70 $\pm$ 3\% respectively. The fraction of \Ha\ luminosity from star-formation and AGN activity is 65 $\pm$ 3\% and 35 $\pm$ 2\% respectively.

The hard ionising radiation field from the AGN produces more \OIII\ than Balmer series lines, as expected. However, we also observe that the AGN is responsible for 35\% of the \Ha\ luminosity of the galaxy and the star-formation is responsible for 30\% of the \OIII\ luminosity - supporting the notion that both starburst and AGN activity are energetically significant in NGC 7130. 

\begin{figure*}
\centerline{\includegraphics[scale=0.75]{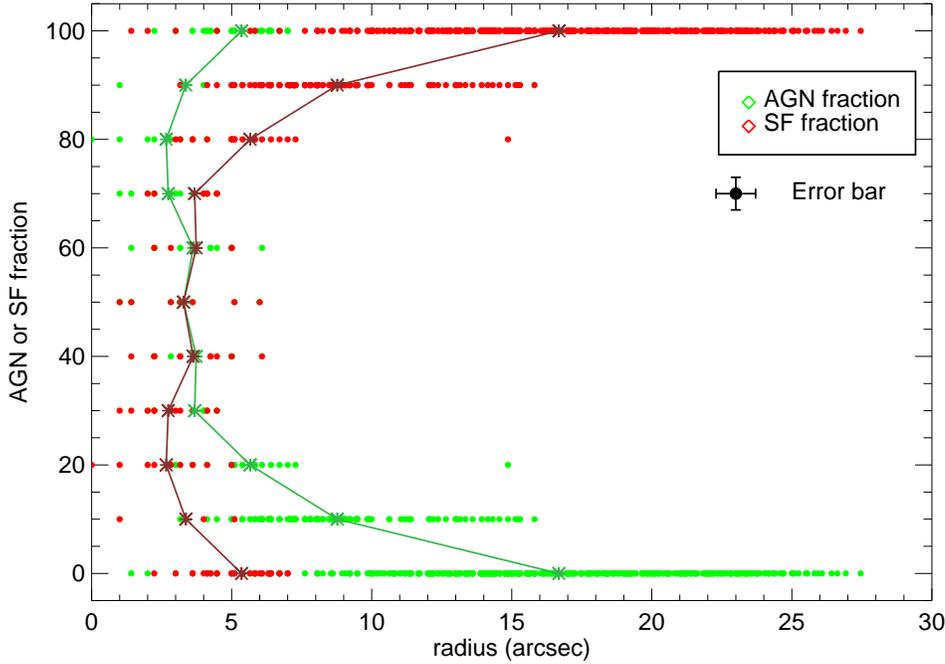}}
\caption{Starburst and AGN fractions as a function of distance from the central AGN. The green points represent the AGN fraction for individual spaxels, whilst the red points represents the starburst fraction. The green and maroon stars indicate the average radii for each AGN fraction and starburst fraction respectively. The error bar on the right hand side of the Figure indicates the nominal error on raw data points at 50\% AGN/SF fraction.}
\label{Fig9}
\end{figure*} 

It is important to note that the AGN fractions indicate the degree of error that is made when using H$\alpha$ as a star formation rate (SFR) indicator in composite galaxies. Use of the \Ha\ luminosity without correction for contribution from an AGN will over-estimate the true SFR of a galaxy (35\% overestimation in the case of NGC 7130). 

\subsection{AGN fraction as a function of radius}
\label{Subsec:Sec55}

In Figure \ref{Fig9}, we show quantitatively how the contributions of the starburst and the AGN change as a function of distance from the centre of NGC 7130. We define the centre of the galaxy to be the centroid of the \OI\ emission, which our analysis from Section \ref{Subsec:Sec41} indicates is the best tracer of AGN activity in the galaxy. The green circles show the AGN fractions of individual spaxels as a function of radius, whilst the red circles trace the relative contribution of the star-formation. The green stars indicate the average distance of spaxels from the centre of the galaxy at each AGN fraction, whilst the maroon stars indicate the average radius of spaxels at each star-forming fraction. The error bar on the right hand side of the Figure indicates the nominal error on raw data points at 50\% AGN/SF fraction. The vertical errors trace the 6\% relative error on the AGN fraction, so that points at higher AGN fraction will have a larger vertical error than that indicated by the nominal error, whilst the vertical error will be lower at lower AGN fractions. The horizontal errors trace the maximum error on the radius due to the PSF of the data. 

The AGN fraction is highest in the centre of the galaxy but drops quickly as the radius increases, with the inverse behaviour observed for the relative contribution of the star-formation. The largest deviation from a monotonic decrease in AGN fraction as a function of radius occurs in the inner-most regions of the galaxy, as the AGN fraction peaks at a non-zero radius. This is likely to be due to the elevated \NIIHa\ and \OIIIHb\ line ratios observed on either side of the \OIHa\ peak in Figure \ref{Fig3}. This highlights that the \OIHa\ vs \OIIIHb\ diagnostic diagram may be a better diagnostic of starburst-AGN mixing when the \OI\ line is detected at a statistically significant level in a large number of spaxels. We will follow this up in a future paper.
		
The distance at which the AGN and the star formation contribute equally (AGN fraction = 50\%) is approximately 3.4 $\pm$ 1.7'', or 1.1 $\pm$ 0.6 kpc.  This is smaller than the extended narrow line region radius derived in Section \ref{Subsec:Sec52} - as expected, since the extended narrow line region is defined such that points outside of it have less than 10\% contribution from the AGN.

\begin{figure}
\centerline{\includegraphics[scale=0.5]{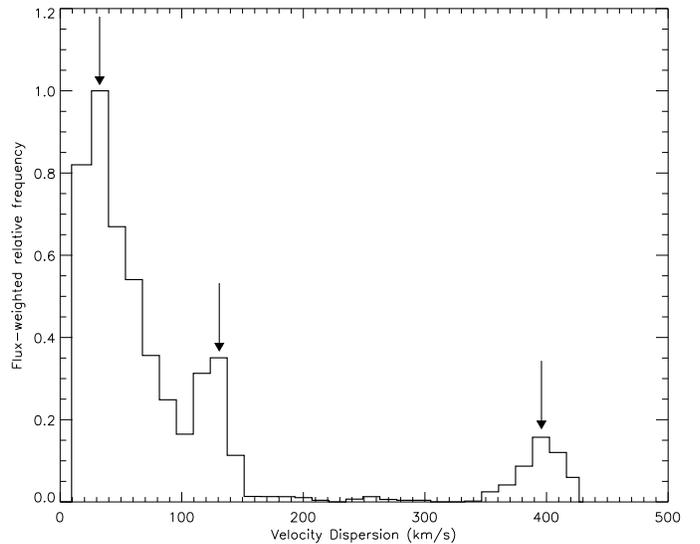}}
\caption{H$\alpha$ flux-weighted velocity dispersion histogram, indicating three peaks at $\sigma \approx \rm 30 \, km s^{-1}$, $\sigma \approx \rm 150 \, km s^{-1}$ and $\sigma \approx \rm 400 \, km s^{-1}$. These peaks are likely to trace the pure star-forming, composite and pure-AGN components respectively.
}\label{Fig10}
\end{figure}

\section{Velocity Field and Velocity Dispersion}
\label{Sec:Sec6}

\begin{figure*}
\centerline{\includegraphics[scale=0.95,clip=true,trim=0 130 0 0]{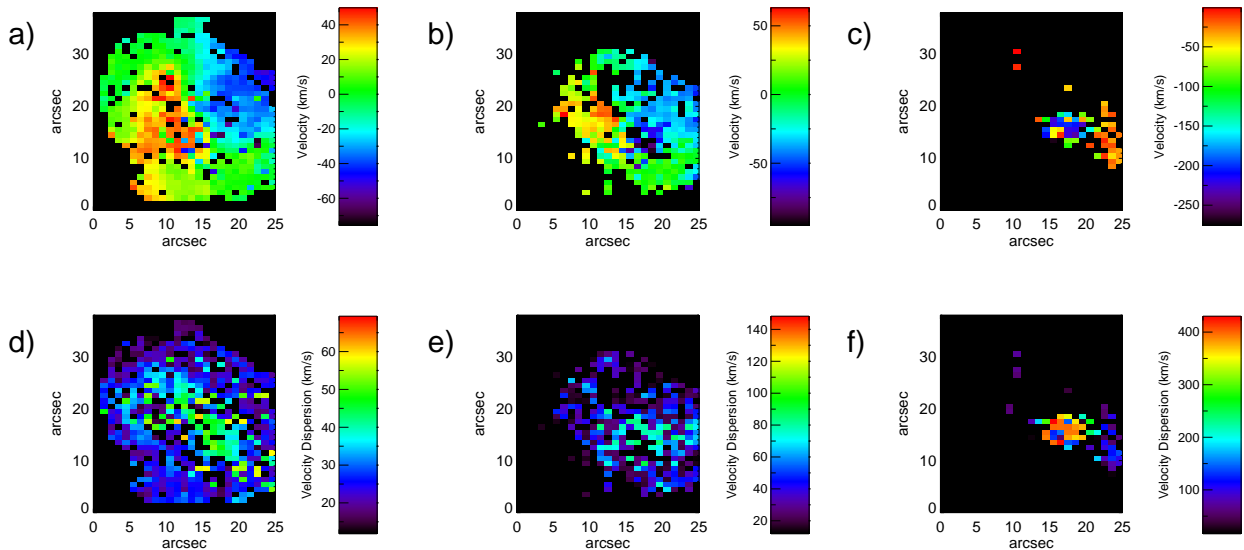}}
\caption{Top, left to right: Maps of velocity field by component (1-3, a to c). The recessional velocities are relative to the overall recessional velocity of the galaxy. Bottom, left to right: maps of velocity dispersion ($\sigma$) by component (1-3, d to f).}
\label{Fig11}
\end{figure*} 

We show the H$\alpha$ flux-weighted velocity dispersion ($\sigma$) distribution for NGC 7130 in Figure \ref{Fig10}. There are three clear peaks, marked by arrows. The first and largest, at $\sigma \approx \rm 30 \, km s^{-1}$, indicates that the optical spectra of NGC 7130 are dominated by low-$\sigma$ \HII\ regions \citep[see e.g.][]{Epinat10}. The other peaks at $\sigma \approx \rm 150 \, km s^{-1}$ and $\sigma \approx 400 \, \rm km s^{-1}$ suggest that there could be distinct velocity components corresponding to more energetic ionisation sources.

We assign three velocity dispersion components: component 1 ($\sigma \leq 80 \rm \, km s^{-1}$), component 2 ($80 \, \rm km s^{-1} < \sigma \leq 250 \, \rm km s^{-1}$), and component 3 ($\sigma > 250 \, \rm km s^{-1}$). We show maps of velocity field as a function of velocity dispersion component in Figure \ref{Fig11} a to c. Maps of velocity dispersion as a function of velocity dispersion component are shown in Figure \ref{Fig11} d to f. 

The velocity field for component 1 is similar to the rotation curve of NGC 7130 across the entire H$\alpha$ disk, and therefore is likely to trace the galaxy's underlying velocity field (see Figure \ref{Fig11} a). The velocity dispersion map shows that $\sigma$ decreases with distance from the centre of the galaxy (see Figure \ref{Fig11} d), peaking at approximately 60$\rm \, km s^{-1}$ and decreasing smoothly to 30$\rm \, km s^{-1}$ at the largest radii. This behaviour is consistent with theoretical and observational views of higher velocity dispersions in Seyfert nuclei \citep[e.g.][]{Stoklasova09}. 

The velocity field and velocity dispersion maps for component 2 show similar features to those of component 1, but over a smaller cross-sectional area (approximately 54\% of the region covered by component 1; see Figures \ref{Fig11} b and e). The higher $\sigma$ in this component suggests that it may trace starburst-AGN mixing.

The third component is detected primarily in spaxels coincident with the nucleus of the \OI/H$\alpha$ emission. The high velocity dispersion (around 400 ${\rm kms}^{-1}$, see Figure \ref{Fig11} f) is consistent with the injection of energy by an active galactic nucleus, and the blue-shifted velocity field (see Figure \ref{Fig11} c) supports the notion of an outflowing wind, as identified by \citet{Veilleux95} and \citet{Levenson02}.

\section{Discussion}
\label{Sec:Sec7}

\begin{figure*}
\centerline{\includegraphics[scale=0.4]{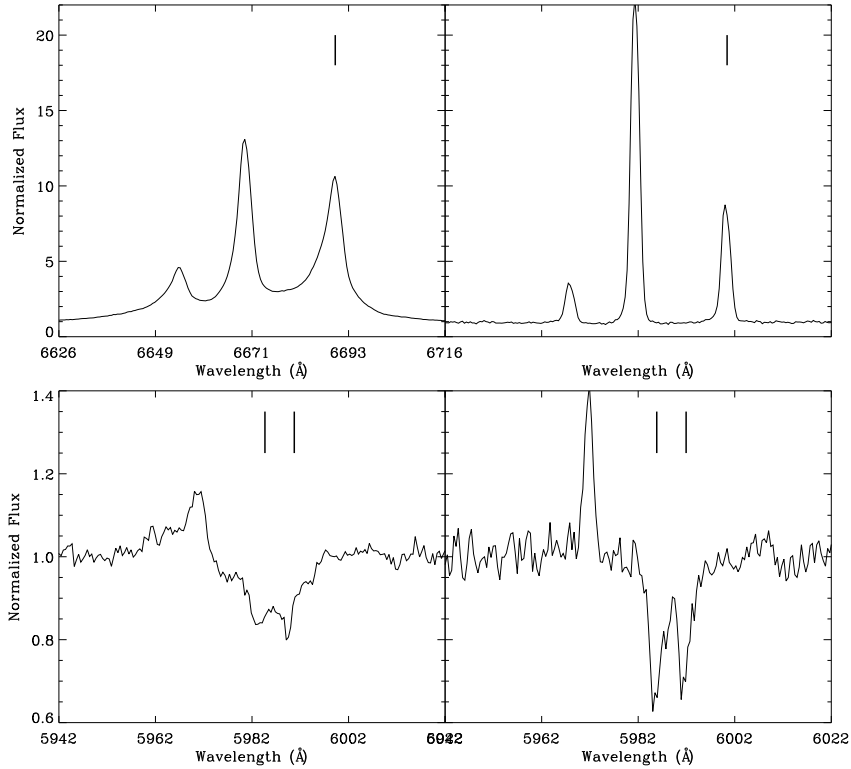}}
\caption{Continuum normalized spectra showing the \mbox{[N II]} + H$\alpha$ (top) and \mbox{He I} + \mbox{Na I D} (bottom) profiles for 5\arcsec apertures from the nucleus (left) and a spiral arm region west of the nucleus (right). The solid vertical lines indicate the rest-frame centroids of \NII\ $\lambda$ 6583 (top) and \mbox{Na I D} $\lambda \, \lambda$ 5890, 5896 (bottom). Strongly blueshifted \mbox{Na I D} and broad emission line profiles in the nuclear region are unambiguous signs of an outflow in the galaxy.}\label{Fig12}
\end{figure*}

Our analysis of line ratio maps, velocity field and velocity dispersion information in NGC 7130 has clearly indicated that this galaxy is a starburst-AGN composite galaxy - a conclusion which is strongly supported by past studies. As a galaxy with a Seyfert 1.9 nucleus, its luminosity might be expected to be dominated by the activity of the AGN. Indeed, the signature of this AGN is clear in the high \NII/H$\alpha$, \OIII/H$\beta$ and \OI/H$\alpha$ ratios and high velocity dispersion, as well as indicators in the IR \citep{Petric11, Pereira-Santaella10, Diaz-Santos10,Stierwalt13} and X-ray \citep{Levenson02}. However, we have demonstrated that this galaxy also shows evidence for strong star-formation - probably triggered by previous merger activity, which remains imprinted on the galaxy's emission signature in the form of its asymmetric optical morphology, streamers and trailing tidal arms. The signature of the star formation is clear from the strong H$\alpha$ emission, the low \NIIHa\ and \OIIIHb\ ratios in the extended emission regions of the galaxy, and the prominent low velocity dispersion component, as well as several other indicators in the IR \citep{GonzalezDelgado01, Contini02}, optical \citep{GonzalezDelgado01}, UV \citep{GonzalezDelgado98}, and X-ray \citep{Levenson02}.

The smooth mixing sequence from the \HII\ to AGN regions on the \NII/H$\alpha$ vs \OIIIHb\ and \SII/H$\alpha$ vs \OIIIHb\ diagnostic diagrams indicates that NGC 7130 contains regions with a range of AGN contributions that depend on distance from the centre of the galaxy. The AGN radiation field is strongest in the nuclear region of the galaxy, and grows weaker as distance from the centre of the galaxy increases until star-formation becomes dominant in the extended emission region of the galaxy. This clean behaviour allows us to identify the boundary of the AGN extended narrow line region, which we find to lie 1.8 $\pm$ 0.8 kpc from the central AGN. The ability to use starburst-AGN mixing to calculate the size of AGN ENLRs could be useful for future investigation of how the size of the ENLR relative to the extent of star formation depends on properties of the AGN. 

We have also presented a technique for numerical calibration of starburst-AGN mixing sequences constructed using spectra from a single galaxy. Our analysis of AGN fractions in NGC 7130 reveals that star-formation dominates the H$\alpha$ luminosity and AGN activity dominates the \OIII\ luminosity (as expected) - however, in both cases the non-dominant energy source is responsible for at least 25\% of the line luminosity - indicating, for example, the degree to which star formation rates calculated from \Ha\ will be overestimated in composite galaxies. Further investigation of the numerical calibration of mixing sequences will assist in the development of techniques to decompose spectra into their star-forming and AGN components (see Paper II for more detailed discussion). 

One source of complication in the emission line analysis for NGC 7130 is the off-nuclear knots of elevated \NII/H$\alpha$ line ratios, seen in Figure \ref{Fig3}. In Section \ref{Sec:Sec4}, we suggested that these may be the result of an outflowing wind shock-heating the interstellar medium. The notion of an outflow is supported in our data by the strong blueshift observed in the third velocity component. In order to investigate this further, we extract profiles of the \NII + H$\alpha$ and \mbox{He I} + \mbox{Na I D} features in 5'' apertures centred on the nucleus and a spiral arm region west of the nucleus (see Figure \ref{Fig12}). The strongly blue-shifted \mbox{Na I D} absorption feature in the nuclear region is an unambiguous sign of outflowing material \citep[e.g.][]{Rupke05}. The nuclear \mbox{He I}, \mbox{[N II]} and H$\alpha$ emission profiles also show significant broadening and blue-shifted tails compared to the clean, narrow line profiles in the spiral arm region. 

\citet{Levenson02} suggest that the outflow is starburst-driven based on the morphology of the extended soft X-ray emission, and \citep{Veilleux95} come to the same conclusion based on the \OIII\ and \Hb\ emission profiles. However, the strong Seyfert-like line ratios seen in emission are more consistent with AGN driven outflows \citep[e.g.][]{Sharp10}. The low spatial resolution of our data makes it difficult to differentiate between a wind driven by circumnuclear starburst or AGN activity. In order to differentiate between starburst and AGN driven winds it is necessary to employ high spatial resolution IFUs on 8-10m class telescopes, preferably with the aid of adaptive optics.

Another possible explanation for the elevated off-nuclear \NIIHa\ line ratios is shock-excitation in an outflowing cocoon. In this scenario, radio jets can interact with the interstellar medium, producing high velocity shocks which provide UV photons to excite the ENLR \citep[see][]{Dopita95, Bicknell97, Bicknell98, Bicknell00}. NGC 7130 has an elongated radio structure which could indicate the presence of radio jets, although this structure has been attributed to shocks or star-formation along the galactic bar \citep{Bransford98}. In order to determine whether out-flowing material or radio jets are responsible for shock-heating the interstellar medium, it is necessary to conduct an analysis of the energetics of the interstellar medium, which is beyond the scope of this paper.

It is also important to consider the impact of metallicity gradients on our interpretation of starburst-AGN mixing. As discussed in Section \ref{Subsec:Sec41}, both the metal abundance of the line-emitting gas and the hardness and concentration of the EUV ionising radiation field from the AGN affect the \NIIHa\ and \OIIIHb\ line ratios. Therefore, for galaxies with metallicity gradients, the variation in \NIIHa\ and \OIIIHb\ as a function of radius is driven by changes in both metal abundance and the AGN fraction. Theoretical starburst-AGN mixing curves constructed using photoionization models suggest that metallicity gradients primarily impact the slope of starburst-AGN mixing sequences. By applying the AGN fraction calculation method we have presented to these theoretical mixing curves, we have determined that the accuracy of our method is not compromised by the presence of abundance gradients when both the abundance gradient and the observed starburst-AGN mixing curve are smooth - confirming that our interpretation of starburst-AGN mixing in NGC 7130 is robust. The impact of abundance gradients on interpretations of starburst-AGN mixing will be discussed in detail in a future paper.

Aside from our major results, this study demonstrates the power of IFU data as a tool which opens the door to many new opportunities in the study of the starburst-AGN connection. Our identification of the outflowing wind in this galaxy clearly highlights the strengths of IFU data in dissecting the power sources of galaxies. The ability to probe the spatial variation of diagnostic line ratios and velocity properties provides us with a much more detailed and far-reaching picture of the physical processes at work within individual galaxies than has ever been derived from single-fibre or long slit spectroscopy. Furthermore, we have shown that more complete and higher resolution data sets allow for more precise numerical calibrations, propelling us into the era of detailed numerical analysis. By combining the results of this study with theoretical investigations and follow-up observational work, we will be able to study star formation rates and metallicity gradients in AGN dominated galaxies as well as pure star-forming galaxies.

\section{Conclusion}
\label{Sec:Sec8}

Using IFU emission line data, we have shown that NGC 7130 is a remarkably clean prototype for starburst-AGN mixing in composite galaxies. By combining line ratio maps and line diagnostic diagrams we have uncovered a clear relationship between the location of points on the starburst-AGN mixing sequence and their distance from the centre of the galaxy. We find that:

\begin{itemize}
\item{The spatial variation of the strong line ratios in NGC 7130 is consistent with a combination of star-formation and AGN activity.}
\item{The composite spaxels, identified using the \NIIHa\ vs \OIIIHb\ diagnostic diagram, form a clear ring around the nuclear region of the galaxy.}
\item{Spaxels which lie further up the mixing sequence on the \NIIHa\ vs \OIIIHb\ diagnostic diagram also lie closer to the centre of the galaxy. When color-coding spaxels according to their distance up the mixing sequence we observe clean rings of gas ionised by different combinations of star formation and AGN activity; indicating that the ionisation state of the gas decreases smoothly with distance from the central AGN.}
\item{We use the clean demarcation between \HII\ region and ENLR gas to calculate the size of the AGN ENLR. We find the radius of the ENLR to be 1.8 $\pm$ 0.8 kpc, consistent with expected values from literature.} 
\item{We present an empirical method for calculating the relative contribution of the star-formation and the AGN activity on a spaxel-by-spaxel basis. We use this to determine that the fraction of \OIII\ luminosity from star-formation and AGN activity is 30 $\pm$ 2\% and 70 $\pm$ 3\% respectively, and that the fraction of \Ha\ luminosity from star-formation and AGN activity is 65 $\pm$ 3\% and 35 $\pm$ 2\% respectively. This highlights the degree of error made when using \Ha\ as a proxy for star-formation rate in composite galaxies.}
\item{Our investigation of the velocity field in NGC 7130 supports the notion of starburst-AGN composite activity, with three separate velocity components identified - one tracing pure star-formation and the underling velocity field of the galaxy, one tracing starburst-AGN mixing over the extended narrow line region of the galaxy, and one tracing the strongly blueshifted and high dispersion nuclear region of the galaxy.}
\item{The emission line and velocity field maps for NGC 7130 also suggest the presence of an outflowing wind in the galaxy; however, our data do not allow us to determine whether this wind is driven by the circumnuclear starburst or the AGN.}
\end{itemize} 

Starburst-AGN mixing has the potential to open the doors to understanding the relationship between star formation and AGN activity, both in individual galaxies and across the entire AGN sequence. In Paper II we will expand this study to a larger sample of galaxies and present starburst-AGN mixing models which provide further support for the observational scenario we have presented. We will also discuss applications of starburst-AGN mixing in more detail, with particular focus on the ability to construct individual spectra for the star-formation and AGN components on a spaxel-by-spaxel basis and the possibility of using observations of starburst-AGN mixing to constrain the conditions of the interstellar medium and the hardness of the AGN ionizing radiation field in starburst-AGN composite galaxies.

\section{Acknowledgements}

We would like to thank the anonymous referee whose comments and suggestions facilitated significant improvement of this manuscript. Kewley \& Dopita acknowledge the support of the Australian Research Council (ARC) through Discovery project DP130103925. This research has made use of the NASA/IPAC Extragalactic Database (NED) which is operated by the Jet Propulsion Laboratory, California Institute of Technology, under contract with the National Aeronautics and Space Administration. We also acknowledge the usage of the HyperLeda Database \href{http://leda.univ-lyon1.fr}{(http://leda.univ-lyon1.fr)} and Aladin Sky Atlas.

\end{document}